# Comparative study between toroidal coordinates and the magnetic dipole field

Esteban Chávez-Alarcón[a] and J. Julio E. Herrera-Velázquez[b]

[a]*Instituto Nacional de Investigaciones Nucleares, Apdo. Postal 18-1027, México, D.F. México*
[b]*Instituto de Ciencias Nucleares, Universidad Nacional Autónoma de México, Apdo. Postal 70-543, Ciudad Universitaria, Delegación Coyoacán, 04511 México, D.F. México*

**Abstract.** There is a similar behaviour between the toroidal coordinates and the dipole magnetic field produced by a circular loop. In this work we evaluate up to what extent the former can be used as a representation of the latter. While the tori in the toroidal coordinates have circular cross sections, those of the circular loop magnetic field are nearly elliptical ovoids, but they are very similar for large aspect ratios. The centres of the latter displace from the axis faster than the former. By making a comparison between tori of similar aspect ratios, we find quantitative criteria to evaluate the accuracy of the approximation.

**Resumen.** Existe cierta semejanza entre las coordenadas toroidales y el campo dipolar magnético producido por una espira circular. En este trabajo se evalúa hasta que punto las primeras pueden ser empleadas como una aproximación del segundo. Mientras que los toros de las coordenadas toroidales tienen secciones transversales circulares, los del campo de una espira circular son ovoides cercanos a elipses. Sin embargo para razones de aspecto grandes, los centros de los segundos se desplazan del eje con mayor rapidez que los primeros. Haciendo una comparación entre toros con razones de aspecto semejantes, se encuentran criterios cuantitativos para evaluar la precisión de la aproximación.

PACS classification: 03.50.De, 02.30.Jr, 41.20.Gz, 52.55.Fa

# 1. Intoduction

Orthogonal systems of coordinates are often chosen according to the symmetries in a given physical problem. Among these, the toroidal coordinates [1, 2] are useful to describe tori with circular cross section. While many problems are nowadays solved numerically, and such tools seem to be falling in disuse, it is still interesting to be able to write down some results in terms of analytical approximations. Mukhovatov and Shafranov, for instance [3], used them in order to draw some important properties of the equilibria in tokamaks. This system allows to separate the Grad-Shafranov equation, which models the magnetohydrodynamic equilibrium of toroidal axisymmetric plasmas.

On the other hand, toroidal coordinates show a significant resemblance with the dipole magnetic field lines produced by a circular current loop, which can be expressed in terms of elliptic integrals, as shown in text books [4, 5]. Since toroidal coordinates are the result of rotating the so called bipolar coordinates around their axis of symmetry, one may be mistakenly lead to believe that they actually describe the field lines of a magnetic dipole. However, the toroidal magnetic field surfaces in this case do not have circular cross sections, as toroidal coordinates do, but rather ovoid shapes. Therefore, the toroidal coordinates will only be a good approximation for large aspect ratios, in which the major radius of the torus is much greater than its minor radius. The purpose of this paper is to compare them in order to evaluate quantitatively the limit in which one system can be approximated by the other, and establish the accuracy of such approximation.

In sections 2 and 3 we shall remind the expressions for the toroidal coordinates and the azimuthal component of the vector potential for a dipole magnetic field, produced by a circular loop current, respectively, in order to establish the notation. In section 4 we shall show how they can be compared, and give quantitative criteria in order to evaluate the accuracy with which toroidal coordinates can be used to approximate the dipole field. Conclusions will be drawn in section 5.

## 2. Toroidal coordinates [2]

Toroidal coordinates ($\eta$, $\xi$, $\varphi$) emerge as a rotation around the $z$ axis of the two dimensional bipolar coordinates Their relationship with the rectangular coordinates ($x$, $y$, $z$) is given by

$$x = \frac{R_o \sinh\eta \cos\varphi}{\cosh\eta - \cos\xi}, \quad y = \frac{R_o \sinh\eta \sin\varphi}{\cosh\eta - \cos\xi}, \quad z = \frac{R_o \sin\xi}{\cosh\eta - \cos\xi}. \tag{1}$$

Figure (1) shows the graphical representation of this system, which we proceed to describe, on the *X-Y* plane:

If we define $\rho^2 = x^2 + y^2$, instead of the first two expressions in eq. (1) we can write

$$\rho = \frac{R_0 \sinh\eta}{\cosh\eta - \cos\xi}, \tag{2}$$

so the set of equations for ($\rho$, $z$) describe the cross sections for any arbitrary toroidal angle $\varphi$.

We are particularly interested in the set of surfaces $\eta$ =ct. ($0 \le \eta \le \infty$), which describe tori with circular cross sections

$$\rho^2 + z^2 + R_o^2 = 2R_o\rho\coth\eta \tag{3}$$

with radii $a=R_o \operatorname{csch}\eta$ (minor radii) centred at $R=R_o \coth\eta$ (major radii) away from the $z$ axis. In the limit $\eta \to 0$ the radii of the circles tend to infinity and the circles degenerate into the $z$ axis, while in the limit $\eta \to \infty$, $R \to R_o$ and $a \to 0$. The aspect ratio of the torii is defined as $A \equiv R/a = \cosh\eta$. It is often more useful to use the inverse aspect ratio $\varepsilon = a/R = \operatorname{sech}\eta$, which is small for thin torii (large values of $\eta$.)

On the other hand, the set of surfaces $\xi =$ ct. ($0 \leq \xi \leq 2\pi$), represent spheres

$$\rho^2 + z^2 - R_o^2 = 2R_o z \cot\xi \quad , \tag{4}$$

centred on the $z$ axis at $R_o \cot\xi$, with radii $R_o|\csc\xi|$. The spheres intersect on the $x, y$ plane at $\rho = R_o$.

The set of half planes defined by $\varphi =$ ct. ($0 \leq \varphi \leq 2\pi$.) represent the azimuthal, or toroidal angle around the $z$ axis.

## 3. The dipole magnetic field produced by the current loop [5]

Finding the magnetic field produced by a circular current loop of radius $R_o$ is an elementary problem of magnetostatics, which is solved by using Ampére's law. Although it is indeed straightforward to compute it in cylindrical coordinates, we start from the expression in spherical coordinates, as obtained in Ref. [5] for the sake of clarity and brevity. Assuming that the loop lies on the $x, y$ plane, centred at the origin, the magnetic field can be written as the curl of the azimuthal component of the vector potential in terms of spherical coordinates $(r, \theta, \varphi)$ as

$$\mathbf{B} = \nabla \times [A_\varphi(r,\varphi)\mathbf{e}_\varphi] \ . \tag{5}$$

For a circular loop carrying a current $I$ we get

$$A_\varphi(r,\theta) = \frac{\mu_o}{4\pi} \frac{4IR_o}{\sqrt{R_o^2 + r^2 + 2R_o r \sin\theta}} \left[ \frac{(2-k^2)K(k) - 2E(k)}{k^2} \right] , \tag{6}$$

where $K(k)$ and $E(k)$ are the complete elliptic integrals

$$K(k) = \int_0^{\frac{\pi}{2}} \frac{d\theta}{\sqrt{1 - k^2 \sin^2\theta}} \ , \quad \text{and} \quad E(k) = \int_0^{\frac{\pi}{2}} \sqrt{(1 - k^2 \sin^2\theta)} d\theta \ , \tag{7}$$

with

$$k = \sqrt{\frac{4R_o r \sin\theta}{R_o^2 + r^2 + 2R_o r \sin\theta}} \ , \tag{8}$$

as shown in Ref. [5].

Since $r^2 = \rho^2 + z^2$, and $\rho = r\sin\theta$, we can rewrite (6) and (8) in terms of the cylindrical coordinates ($\rho, \varphi, z$) as

$$A_\varphi(\rho,z) = \frac{\mu_o I}{2\pi} \frac{\sqrt{(R_o + \rho)^2 + z^2}}{\rho} \left\{ \frac{R_o^2 + \rho^2 + z^2}{(R_o + \rho)^2 + z^2} K(k) - E(k) \right\} , \tag{6a}$$

$$k = \sqrt{\frac{4R_0\rho}{(R_0+\rho)^2 + z^2}} \quad . \tag{8a}$$

It is useful to write $A_\varphi$ in terms of the stream function $\psi$ as $A_\varphi = \rho^{-1}\psi$, since $\psi$ is a measure of the magnetic flux $\int \mathbf{B} \cdot \mathbf{dS}$, where the surface of integration would be a ring between $R_o$ and $\rho$, lying on the plane $z = 0$. Since $\nabla\varphi = \rho^{-1}\mathbf{e}_\varphi$, the magnetic field will be given by

$$\mathbf{B}(\rho,z) = \nabla \times \left(\frac{\psi(\rho,z)}{\rho}\mathbf{e}_\varphi\right) = \nabla \times (\psi(\rho,z)\nabla\varphi) = \nabla\psi(\rho,z) \times \nabla\varphi \quad . \tag{9}$$

Therefore, making use of (6a),

$$\psi(\rho,z) = \frac{\mu_o I}{2\pi}\sqrt{(R_o+\rho)^2 + z^2}\left\{\frac{R_o^2 + \rho^2 + z^2}{(R_o+\rho)^2 + z^2}K(k) - E(k)\right\} \quad . \tag{10}$$

**4. Comparison between toroidal coordinates and the magnetic flux function**

Since $\mathbf{B} \cdot \nabla\psi = 0$, as can be seen from (9), $\psi$ can be used as a flux function, describing tori of nearly elliptic ovoidal cross section, similar to those represented by $\eta = \sec h(a/R) = ct.$, in the case of the toroidal coordinates.

In Figure 2 some of the circular cross-sections from the toroidal coordinates are plotted as dashed circles, along with surfaces of constant $\psi$. Since the latter are not circular, but resemble ellipses, we need to choose an adequate way to compare them. For this purpose, we define $\varepsilon_\eta = a/R = a/(R_o \coth\eta)$, and $\varepsilon_\psi$ as the ratio of the semiaxis of the ellipses defined by $\psi$ = ct. and their centres, $R_\psi$, away from the *z* axis on the *x, y* plane. The three cases compared in Fig. 2, with a dot marking their intersection, are shown in bold in the table. We have furthermore normalised the coordinates $\rho$ and *z* to $R_o$, keeping in mind that *R* will be different for each circle. As expected, surfaces of constant $\psi$ and $\eta$ with similar values of $\varepsilon_\eta$ and $\varepsilon_\psi$ tend to match when they are small, up to $\varepsilon_\eta$, $\varepsilon_\psi \sim .04$. However, as they grow, the major radius of the toroidal coordinate tori shifts outwards to $R=R_o \coth\eta$, while $R_\psi$ shifts faster, as can be seen in Figure 3. The deformation of the $\psi$ surfaces can also be observed in Fig. 2.

It can also be seen that the inverse aspect ratio grows more slowly for the family of curves $\psi$. In order to have a better understanding of this, we need a way to compare directly $\psi$ and $\eta$. Choosing the $\varepsilon_\eta$, $\varepsilon_\psi = .02$ pair of surfaces as a reference, we normalise the values of $\psi$ and $\eta$ to 1. The normalised functions $\eta_n$ and $\psi_n$ are plotted in Figure 4, where the common parameter for each point is the value of $\varepsilon_\eta \sim \varepsilon_\psi$. It is seen that their dependence is linear, for small values of $\varepsilon$. By fitting them, it is found that they can be related as $\psi_n = 0.9878 \eta_n^{1.14}$.

The accuracy with which the magnetic field of the circular loop can be approximated by the toroidal coordinates, can be appreciated in Figure 5. Table 1 shows eight pairs of surfaces, chosen in such a way that $\varepsilon_\eta$ and $\varepsilon_\psi$ have similar values. The way in which $\eta_n$ and $\psi_n$ differ is shown in terms of percentage.

## 5. Conclusions

A comparison between tori from toroidal coordinates and the magnetic field produced by a circular current loop shows that they can be used interchangeably with a rather good approximation up to inverse aspect ratios in the order of 0.04 ($\varepsilon^{-1}=25$), where they differ by 2.5%. While the normalised values of tori with comparable aspect ratios differ only up to 15.5%, when the inverse aspect ratio grows up to 0.3 ($\varepsilon^{-1}=3.3$), the main problem with the approximation is that the shift in the major radii of both systems differ significantly, as seen in Figure 4.

**Acknowledgements**

This paper was partially supported by CONACyT grant G133036 and DGAPA-UNAM grant IN120409-3.

| $\varepsilon_\eta$ | $\eta_n$ | $\varepsilon_\psi$ | $\psi_n$ | $\dfrac{\eta_n - \psi_n}{\psi_n} \times 100\%$ |
|---|---|---|---|---|
| 0.02067 | 1 | 0.02065 | 1 | 0 |
| **0.04001** | **0.85547** | **0.0399** | **0.8347** | **2.5** |
| 0.09325 | 0.67004 | 0.09227 | 0.62667 | 6.9 |
| **0.14173** | **0.57783** | **0.13891** | **0.52678** | **9.7** |
| 0.1868 | 0.51661 | 0.18127 | 0.46236 | 11.7 |
| 0.22903 | 0.47104 | 0.22009 | 0.41579 | 13.3 |
| **0.26877** | **0.43491** | **0.25587** | **0.37985** | **14.5** |
| 0.30624 | 0.40511 | 0.28901 | 0.35083 | 15.5 |

**Table 1.** Using the set of surfaces with $\varepsilon_\eta =.02067$, $\varepsilon_\psi = 0.02065$ in order to normalise $\eta$ and $\psi$ to 1, we give the values for other pairs of surfaces such that $\varepsilon_\eta$ and $\varepsilon_\psi$ are close, up to two significant figures, and plot them in Figs. 2 and 3. The cases shown in bold are shown in Fig. 1.

**Figure Captions**

**Figure 1.** Graphical representation of the toroidal coordinates system (1) on the *X-Z* plane ($\varphi = 0$). The toroids are generated by rotating the graph around the *Z* axis.

**Figure 2.** Comparison between toroidal coordinates (dashed) and curves of constant values of $\psi$ (solid), where the inverse aspect ratio of the toroidal case, $\varepsilon_\eta = a/R_o = \mathrm{sech}\,\eta$, is close to the inverse aspect ratio of the loop current magnetic field $\varepsilon_\psi$, defined as the ratio of the inner major radius of the ellipse and its major radius. The dots mark the point of intersection of the matching pairs $\varepsilon_{\eta=}.04$, $\varepsilon_{\psi=}0.0399$; $\varepsilon_{\eta=}.14173$, $\varepsilon_{\psi=}0.13891$, and $\varepsilon_{\eta=}.26877$, $\varepsilon_{\psi=}0.25587$.

**Figure 3.** Displacement of the major radius, normalised to the position of the current loop, $R_o$, as a function of the inverse aspect ration $\varepsilon$. $R_\eta = R_o \coth\eta$ shows the evolution of the displacement for toroidal coordinates, while $R_\psi$ shows the one of the toroids that arise from the flux coordinate $\psi$ obtained from the circular current loop.

**Figure 4.** Comparison between $\psi$ and $\eta$ matching pairs of coordinates with common values of $\varepsilon$. In order to have a proper comparison, we plot $\psi_n$ and $\eta_n$, which have been normalised to 1 for the pair of surfaces with $\varepsilon=.02$. Their relationship is $\psi_n = 0.9878\eta^{1.14}$.

**Figure 5.** Normalised values of $\psi_n$ and $\eta_n$ vs. $\varepsilon = \varepsilon_\psi \sim \varepsilon_\eta$. Predictably, the difference grows with $\varepsilon$ The values of $\psi_n$ are represented by dots (red in the electronic version), while those of $\eta_n$ are shown by squares. Some of the data used are given in Table 1.